\documentclass[a4paper]{PoS}
\usepackage{amsmath}
\usepackage{amssymb}
\usepackage{epsfig}
\usepackage{euscript}
\def\mathswitchr#1{\relax\ifmmode{\mathrm{#1}}\else$\mathrm{#1}$\fi}

\newcommand {\pslash}{\hbox{$\not\hbox{\kern-2.3pt $p$}$}}

\usepackage{cite}

\usepackage{epic}
\def\alf1{ {\alpha\over\pi} }

\title{An Estimate of Lambda in Resummed Quantum Gravity in the Context of Asymptotic Safety and Planck Scale Cosmology: Constraints on SUSY GUTS}

\ShortTitle{An Estimate of Lambda in Resummed Quantum Gravity in the Context of Asymptotic Safety and Planck Scale Cosmology: Constraints on SUSY GUTS}

\author{\speaker{B.F.L. Ward}\\
        Baylor University\\
        E-mail: \email{bfl\_ward@baylor.edu}}

\abstract{We use the amplitude-based resummation of Feynman`s formulation of 
Einstein`s theory to arrive at a UV finite approach to quantum gravity. We show
that we recover the UV fixed point recently claimed by the exact field-space 
renormalization group approach. We use our approach in the context of the 
attendant Planck scale cosmology formulation of Bonanno and Reuter to estimate 
the value of the cosmological constant as $\rho_\Lambda=(0.0024 eV)^4$. We show 
that the closeness of this estimate to experiment constrains susy GUT models.}

\FullConference{36th International Conference on High Energy Physics\\
                 4-11 July 2012\\
                 Melbourne, Australia\\
                BU-HEPP-12-06, Nov., 2012}

\begin{document}
\baselineskip=10pt
\section{\bf Introduction}\label{intro}
\par
In Ref.~\cite{wein1}, Weinberg suggested that the general theory of relativity may be asymptotically safe, with an S-matrix
that depends on only a finite number of observable parameters, due to
the presence of a non-trivial UV fixed point, with a finite dimensional critical surface
in the UV limit. 
Strong evidence has been 
calculated~\cite{reutera,laut,reuterb,reuter3,litim,perc}
using Wilsonian~\cite{kgw} field-space exact renormalization 
group methods to support
Weinberg's asymptotic safety hypothesis for the Einstein-Hilbert theory.
A parallel but independent development~\cite{bw1,bw2,bw2a}, has shown~\cite{bw2i} that the extension of the amplitude-based, exact resummation theory of Ref.~\cite{yfs} to the Einstein-Hilbert theory leads to UV-fixed-point behavior for the dimensionless
gravitational and cosmological constants and to a resummed theory,
resummed quantum gravity, that
is actually UV finite. 
More evidence for Weinberg's asymptotic safety behavior has been calculated using causal dynamical triangulated lattice methods in Ref.~\cite{ambj}\footnote{We also note that the model in Ref.~\cite{horva} realizes many aspects
of the effective field theory implied by the anomalous dimension of 2 at the
UV-fixed point but it does so at the expense of violating Lorentz invariance.}.
\par
The results in Refs.~\cite{reutera,laut,reuterb,reuter3,litim,perc}, while impressive, involve cut-offs and some dependence on gauge parameters
which remain in the results
to varying degrees even for products such as that for the UV limits of the 
dimensionless gravitational and cosmological constants.  
Accordingly, we refer to the approach in 
Refs.~\cite{reutera,laut,reuterb,reuter3,litim,perc} as the 
'phenomenological' asymptotic safety approach.
The above noted 
dependencies are mild enough that the non-Gaussian UV 
fixed point found in these references is probably a physical result. 
But, the results cannot be considered final until a 
rigorously cut-off independent and gauge invariant calculation corroborates 
them. Our approach offers such a possibility, as our results are both gauge invariant and cut-off independent. 
The results from Refs.~\cite{ambj} involve 
lattice constant-type artifact 
issues -- to be considered final they too need to be corroborated by a rigorous calculation 
without such issues. Again, our approach offers a possible 
answer. The stage is therefore prepared for us to try to make contact with experiment.\par
Accordingly, we note that, in Refs.~\cite{reuter1,reuter2}, 
it has been argued that the attendant phenomenological
asymptotic safety approach in 
Refs.~\cite{reutera,laut,reuterb,reuter3,litim,perc} 
to quantum gravity
may indeed provide a realization\footnote{The attendant 
choice of the scale $k\sim 1/t$ used in Refs.~\cite{reuter1,reuter2} 
was also proposed in Ref.~\cite{sola1}.} of the successful
inflationary model~\cite{guth,linde} of cosmology
without the need of the inflaton scalar field: the attendant UV fixed point solution
allows one to develop Planck scale cosmology that joins smoothly onto
the standard Friedmann-Walker-Robertson classical descriptions.
One arrives at a quantum mechanical 
solution to the horizon, flatness, entropy
and scale free spectrum problems. In Ref.~\cite{bw2i}, we have shown
that, in the new
resummed theory~\cite{bw1,bw2,bw2a} of quantum gravity, 
we reproduce the properties as used in Refs.~\cite{reuter1,reuter2} 
for the UV fixed point of quantum gravity with
``first principles''
predictions for the fixed point values of
the respective dimensionless gravitational and cosmological constants. 
Here, we carry the analysis forward to arrive at an estimate
for the observed cosmological constant $\Lambda$ in the
context of the Planck scale cosmology of Refs.~\cite{reuter1,reuter2}.
We comment on the reliability of the result, as it will be seen
already to be relatively close to the observed value~\cite{cosm1,pdg2008}.
While we obviously do not want to overdo the closeness to the experimental value, we feel that this again gives, at the least, some more credibility to the new resummed theory as well as to the methods in Refs.~\cite{reutera,laut,reuterb,reuter3,litim,perc,ambj}. We also show how
the closeness of our estimate 
to the observed value would constrain SUSY GUT models
when this closeness is put on a more firm basis. More reflections on
such matters will be taken up
elsewhere~\cite{elswh}.\par
The discussion is organized as follows. We start by recapitulating in the next section 
the Planck scale cosmology presented phenomenologically
in Refs.~\cite{reuter1,reuter2}. 
We then review briefly in Section 3 our results in
Ref.~\cite{bw2i} for the dimensionless gravitational and cosmological constants
at the UV fixed point. In Section 4, we 
combine the Planck scale cosmology 
scenario in Refs.~\cite{reuter1,reuter2} with our results to estimate 
the observed value of 
the cosmological constant $\Lambda$ and we use it to constrain
SUSY GUTs. 
\par
\section{\bf Planck Scale Cosmology}
We recall the Einstein-Hilbert 
theory
\begin{equation}
{\cal L}(x) = \frac{1}{2\kappa^2}\sqrt{-g}\left( R -2\Lambda\right)
%            + \sqrt{-g} L^{\cal G}_{SM}(x)
\label{lgwrld1a}
\end{equation} 
where $R$ is the curvature scalar, $g$ is the determinant of the metric
of space-time $g_{\mu\nu}$, $\Lambda$ is the cosmological
constant and $\kappa=\sqrt{8\pi G_N}$ for Newton's constant
$G_N$. The authors in Ref.~\cite{reuter1,reuter2},
using the phenomenological exact renormalization group
for the Wilsonian~\cite{kgw} coarse grained effective 
average action in field space,  
have argued that
the attendant running Newton constant $G_N(k)$ and running 
cosmological constant
$\Lambda(k)$ approach UV fixed points as $k$ goes to infinity
in the deep Euclidean regime in the sense that 
$k^2G_N(k)\rightarrow g_*,\; \Lambda(k)\rightarrow \lambda_*k^2$
for $k\rightarrow \infty$ in the Euclidean regime.\par
The contact with cosmology proceeds as follows. Using a phenomenological
connection between the momentum scale $k$ characterizing the coarseness
of the Wilsonian graininess of the average effective action and the
cosmological time $t$, the authors
in Refs.~\cite{reuter1,reuter2} show that the standard cosmological
equations admit of the following extension:
$
(\frac{\dot{a}}{a})^2+\frac{K}{a^2}=\frac{1}{3}\Lambda+\frac{8\pi}{3}G_N\rho,\;
\dot{\rho}+3(1+\omega)\frac{\dot{a}}{a}\rho=0,\;
\dot{\Lambda}+8\pi\rho\dot{G_N}=0,\;
G_N(t)=G_N(k(t)),\;
\Lambda(t)=\Lambda(k(t))
$
in a standard notation for the density $\rho$ and scale factor $a(t)$
with the Robertson-Walker metric representation as
\begin{equation}
ds^2=dt^2-a(t)^2\left(\frac{dr^2}{1-Kr^2}+r^2(d\theta^2+\sin^2\theta d\phi^2)\right)
\label{metric1}
\end{equation}
so that $K=0,1,-1$ correspond respectively to flat, spherical and
pseudo-spherical 3-spaces for constant time t.  The equation of state
is taken as 
$ 
p(t)=\omega \rho(t),
$
where $p$ is the pressure. 
The attendant functional relationship between the respective
momentum scale $k$ and the cosmological time $t$ is determined
phenomenologically via
$
k(t)=\frac{\xi}{t}
$
for some positive constant $\xi$ determined
from constraints on
physically observable predictions.\par
Using the UV fixed points as discussed above for $k^2G_N(k)\equiv g_*$ and
$\Lambda(k)/k^2\equiv \lambda_*$ obtained from their phenomenological, exact renormalization
group (asymptotic safety) 
analysis, the authors in Refs.~\cite{reuter1,reuter2}
show that the system given above admits, for $K=0$,
a solution in the Planck regime where $0\le t\le t_{\text{class}}$, with
$t_{\text{class}}$ a ``few'' times the Planck time $t_{Pl}$, which joins
smoothly onto a solution in the classical regime, $t>t_{\text{class}}$,
which coincides with standard Friedmann-Robertson-Walker phenomenology
but with the horizon, flatness, scale free Harrison-Zeldovich spectrum,
and entropy problems all solved purely by Planck scale quantum physics.\par
While the dependencies of
the fixed-point results $g_*,\lambda_*$ on the cut-offs
used in the Wilsonian coarse-graining procedure, for example,
make the phenomenological nature of the analyses in Refs.~\cite{reuter1,reuter2} manifest, we note that 
the key properties of $g_*,\; \lambda_*$ used for these analyses 
are that the two UV limits are both positive and that the product 
$g_*\lambda_*$ is only mildly cut-off/threshold function dependent.
Here, we review the predictions in Refs.~\cite{bw2i} for these
UV limits as implied by resummed quantum gravity(RQG) theory as presented in
~\cite{bw1,bw2,bw2a}
and show how to use them to predict the current value of $\Lambda$.
For completeness, we start the next section
with a brief review of the basic principles of RQG theory. 
%In this way, we put the arguments in Refs.~\cite{reuter1,reuter2} on a more rigorous theoretical basis.\par 
\par
%\section{Elements of Resummed Quantum Gravity and}
\section{\bf $g_*$ and $\lambda_*$ in Resummed Quantum  Gravity}
We start with the prediction for $g_*$, which we already presented in Refs.~\cite{bw1,bw2,bw2a,bw2i}. Given that
the theory we use is not very familiar, we recapitulate
the main steps in the calculation.
\par
As the graviton couples to an elementary particle 
in the infrared regime which we shall
resum independently of the particle's spin~\cite{wein-qft}, 
we may use a scalar
field to develop the required calculational framework, which we then extend
to spinning particles straightforwardly.  
We follow Feynman in Refs.~\cite{rpf1,rpf2} 
and start with the Lagrangian density for
the basic scalar-graviton system:
\begin{equation}
\begin{split}
{\cal L}(x) &= -\frac{1}{2\kappa^2} R \sqrt{-g}
            + \frac{1}{2}\left(g^{\mu\nu}\partial_\mu\varphi\partial_\nu\varphi - m_o^2\varphi^2\right)\sqrt{-g}\\
            &= \quad \frac{1}{2}\left\{ h^{\mu\nu,\lambda}\bar h_{\mu\nu,\lambda} - 2\eta^{\mu\mu'}\eta^{\lambda\lambda'}\bar{h}_{\mu_\lambda,\lambda'}\eta^{\sigma\sigma'}\bar{h}_{\mu'\sigma,\sigma'} \right\}\\
            & \qquad + \frac{1}{2}\left\{\varphi_{,\mu}\varphi^{,\mu}-m_o^2\varphi^2 \right\} -\kappa {h}^{\mu\nu}\left[\overline{\varphi_{,\mu}\varphi_{,\nu}}+\frac{1}{2}m_o^2\varphi^2\eta_{\mu\nu}\right]\\
            & \quad - \kappa^2 \left[ \frac{1}{2}h_{\lambda\rho}\bar{h}^{\rho\lambda}\left( \varphi_{,\mu}\varphi^{,\mu} - m_o^2\varphi^2 \right) - 2\eta_{\rho\rho'}h^{\mu\rho}\bar{h}^{\rho'\nu}\varphi_{,\mu}\varphi_{,\nu}\right] + \cdots \\
\end{split}
\label{eq1-1}
\end{equation}
Here,
$\varphi(x)$ can be identified as the physical Higgs field as
our representative scalar field for matter,
$\varphi(x)_{,\mu}\equiv \partial_\mu\varphi(x)$,
and $g_{\mu\nu}(x)=\eta_{\mu\nu}+2\kappa h_{\mu\nu}(x)$
where we follow Feynman and expand about Minkowski space
so that $\eta_{\mu\nu}=diag\{1,-1,-1,-1\}$.
We have introduced Feynman's notation
$\bar y_{\mu\nu}\equiv \frac{1}{2}\left(y_{\mu\nu}+y_{\nu\mu}-\eta_{\mu\nu}{y_\rho}^\rho\right)$ for any tensor $y_{\mu\nu}$\footnote{Our conventions for raising and lowering indices in the 
second line of (\ref{eq1-1}) are the same as those
in Ref.~\cite{rpf2}.}.
The bare(renormalized) scalar boson mass here is $m_o$($m$) 
and we set presently the small
observed~\cite{cosm1,pdg2008} value of the cosmological constant
to zero so that our quantum graviton, $h_{\mu\nu}$, has zero rest mass.
We return to the latter point, however, when we discuss phenomenology.
%Here, our normalizations are such that $\kappa=\sqrt{8\pi G_N}$
%where $G_N$ is Newton's constant.
Feynman~\cite{rpf1,rpf2} has essentially worked out the Feynman rules for (\ref{eq1-1}), including the rule for the famous
Feynman-Faddeev-Popov~\cite{rpf1,ffp1a,ffp1b} ghost contribution required 
for unitarity with the fixing of the gauge
(we use the gauge of Feynman in Ref.~\cite{rpf1},
$\partial^\mu \bar h_{\nu\mu}=0$).
For this material we refer to Refs.~\cite{rpf1,rpf2}. 
We turn now directly to the quantum loop corrections
in the theory in (\ref{eq1-1}).
\par
\begin{figure}
\begin{center}
\epsfig{file=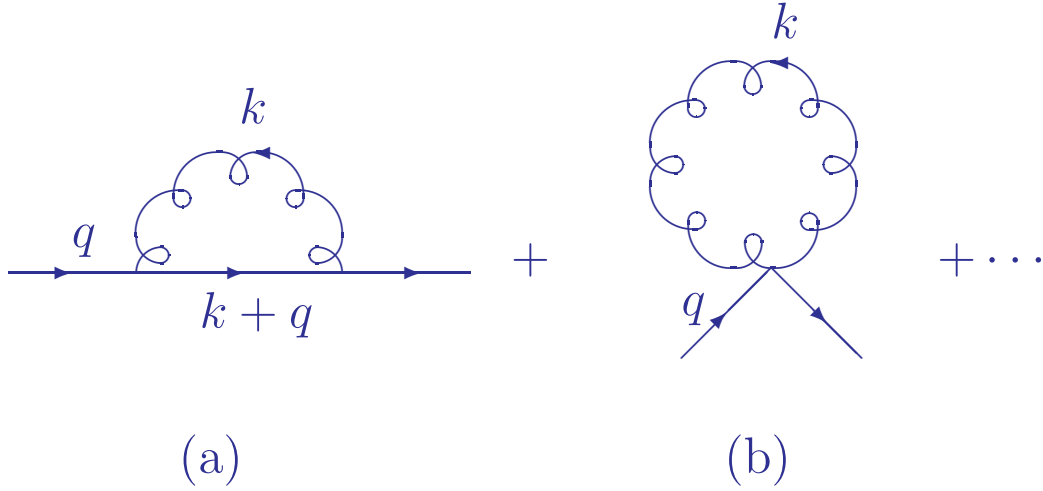,width=140mm}
\end{center}
\caption{\baselineskip=7mm     Graviton loop contributions to the
scalar propagator. $q$ is the 4-momentum of the scalar.}
\label{fig1}
\end{figure}

Referring to Fig.~\ref{fig1}, 
we have shown in Refs.~\cite{bw1,bw2,bw2a} that the large virtual IR effects
in the respective loop integrals for 
the scalar propagator in quantum general relativity 
can be resummed to the {\em exact} result
$
i\Delta'_F(k)=\frac{i}{k^2-m^2-\Sigma_s(k)+i\epsilon}
=  \frac{ie^{B''_g(k)}}{k^2-m^2-\Sigma'_s+i\epsilon}
\equiv i\Delta'_F(k)|_{\text{resummed}}
%&=\frac{i}{k^2-m^2-\Sigma_s(k)+i\epsilon}
$
for{\small ~~~($\Delta =k^2 - m^2$)
\begin{equation}
\begin{split} 
B''_g(k)&= -2i\kappa^2k^4\frac{\int d^4\ell}{16\pi^4}\frac{1}{\ell^2-\lambda^2+i\epsilon}\\
&\qquad\frac{1}{(\ell^2+2\ell k+\Delta +i\epsilon)^2}\\
&=\frac{\kappa^2|k^2|}{8\pi^2}\ln\left(\frac{m^2}{m^2+|k^2|}\right),       
\end{split}
\label{yfs1} 
\end{equation}}
where the latter form holds for the UV(deep Euclidean) regime, 
so that $\Delta'_F(k)|_{\text{resummed}}$ 
falls faster than any power of $|k^2|$ -- by Wick rotation, the identification
$-|k^2|\equiv k^2$ in the deep Euclidean regime gives 
immediate analytic continuation to the result in the last line of (\ref{yfs1})
when the usual $-i\epsilon,\; \epsilon\downarrow 0,$ is appended to $m^2$. An analogous result~\cite{bw1} holds
for m=0. Here, $-i\Sigma_s(k)$ is the 1PI scalar self-energy function
so that $i\Delta'_F(k)$ is the exact scalar propagator. As $\Sigma'_s$ starts in ${\cal O}(\kappa^2)$,
we may drop it in calculating one-loop effects. 
When the respective analogs of $i\Delta'_F(k)|_{\text{resummed}}$\footnote{These follow from
the observation~\cite{bw1,wein-qft} that the IR limit of the coupling of the graviton to a particle is independent of its spin.} are used for the
elementary particles, one-loop 
corrections are finite. In fact, the use of
our resummed propagators renders all quantum 
gravity loops UV finite~\cite{bw1,bw2,bw2a}. It is this attendant representation
of the quantum theory of general relativity 
that we have called resummed quantum gravity (RQG).
\par
Indeed,
when we use our resummed propagator results, 
as extended to all the particles
in the SM Lagrangian and to the graviton itself, working now with the
complete theory
$
{\cal L}(x) = \frac{1}{2\kappa^2}\sqrt{-g} \left(R-2\Lambda\right)
            + \sqrt{-g} L^{\cal G}_{SM}(x)
$
where $L^{\cal G}_{SM}(x)$ is SM Lagrangian written in diffeomorphism
invariant form as explained in Refs.~\cite{bw1,bw2a}, we show in the Refs.~\cite{bw1,bw2,bw2a} that the denominator for the propagation of transverse-traceless
modes of the graviton becomes ($M_{Pl}$ is the Planck mass)
$
q^2+\Sigma^T(q^2)+i\epsilon\cong q^2-q^4\frac{c_{2,eff}}{360\pi M_{Pl}^2},
$
where we have defined
$
c_{2,eff}=\sum_{\text{SM particles j}}n_jI_2(\lambda_c(j))
         \cong 2.56\times 10^4
$
with $I_2$ defined~\cite{bw1,bw2,bw2a}
by
$
%\begin{split}
%c_1=I_1(\lambda_c)&=\int^{\infty}_0dx x^3(1+x)^{-3-\lambda_c x}\\
I_2(\lambda_c) =\int^{\infty}_0dx x^3(1+x)^{-4-\lambda_c x}
%\end{split}
$
and with $\lambda_c(j)=\frac{2m_j^2}{\pi M_{Pl}^2}$ and~\cite{bw1,bw2,bw2a}
$n_j$ equal to the number of effective degrees of particle $j$. 
The details of the derivation of the
numerical value of $c_{2,eff}$ are given in Refs.~\cite{bw1}.
These results allow us to identify (we use $G_N$ for $G_N(0)$) 
$
G_N(k)=G_N/(1+\frac{c_{2,eff}k^2}{360\pi M_{Pl}^2})
$
and to compute the UV limit $g_*$ as
$
g_*=\lim_{k^2\rightarrow \infty}k^2G_N(k^2)=\frac{360\pi}{c_{2,eff}}\cong 0.0442.$
%We stress that this result has no threshold/cut-off effects in it.
%It is a pure property of the known world.
\par
For the prediction for $\lambda_*$, we use the Euler-Lagrange
equations to get Einstein's equation as 
\begin{equation}
G_{\mu\nu}+\Lambda g_{\mu\nu}=-\kappa^2 T_{\mu\nu}
\label{eineq1}
\end{equation}
in a standard notation where $G_{\mu\nu}=R_{\mu\nu}-\frac{1}{2}Rg_{\mu\nu}$,
$R_{\mu\nu}$ is the contracted Riemann tensor, and
$T_{\mu\nu}$ is the energy-momentum tensor. Working then with
the representation $g_{\mu\nu}=\eta_{\mu\nu}+2\kappa h_{\mu\nu}$
for the flat Minkowski metric $\eta_{\mu\nu}=\text{diag}(1,-1,-1,-1)$
we see that to isolate $\Lambda$ in Einstein's 
equation (\ref{eineq1}) we may evaluate
its VEV(vacuum expectation value of both sides). 
On doing this as described in Refs.~\cite{bw1}, we see that 
%For any bosonic quantum field $\varphi$ we use
%the point-splitting definition\footnote{We need to stress that this is a definition of convenience and is {\em not} a regularization because the integral which we calculate in (\ref{lambscalar}) below it is UV finite with exponential damping in the UV. The definition is robust, the direction of approach to the origin can be chosen arbitrarily, and when its vacuum expectation value is taken it may be replaced with the standard path integral Feynman rule for the tadpole loop that it most certainly is to give the same result.}  (here, :~~: denotes normal ordering as usual)
%\begin{equation}
%\begin{split}
%\varphi(0)\varphi(0)&=\lim_{\epsilon\rightarrow 0}\varphi(\epsilon)\varphi(0)\cr
%&=\lim_{\epsilon\rightarrow 0} T(\varphi(\epsilon)\varphi(0))\cr
%&=\lim_{\epsilon\rightarrow 0}\{ :(\varphi(\epsilon)\varphi(0)): + <0|T(\varphi(\epsilon)\varphi(0))|0>\}\cr
%\end{split}
%\end{equation}
%where the limit $\epsilon\equiv(\epsilon,\vec{0})\rightarrow (0,0,0,0)\equiv 0$
%is taken from a time-like direction respectively. Thus, 
a scalar makes the contribution to $\Lambda$ given by\footnote{We note the
use here in the integrand of $2k_0^2$ rather than the $2(\vec{k}^2+m^2)$ in Ref.~\cite{bw2i}, to be
consistent with $\omega=-1$~\cite{zeld} for the vacuum stress-energy tensor.}
\begin{equation}
\Lambda_s=-8\pi G_N\frac{\int d^4k}{2(2\pi)^4}\frac{(2k_0^2)e^{-\lambda_c(k^2/(2m^2))\ln(k^2/m^2+1)}}{k^2+m^2}
\cong -8\pi G_N[\frac{1}{G_N^{2}64\rho^2}],
\label{lambscalar}
\end{equation} 
where $\rho=\ln\frac{2}{\lambda_c}$ and we have used the calculus
of Refs.~\cite{bw1,bw2,bw2a}.
%as recapitulated here in Appendices 2,3. 
The standard methods~\cite{bw1} 
%equal-time (anti-)commutation 
%relations algebra realizations
then show that a Dirac fermion contributes $-4$ times $\Lambda_s$ to
$\Lambda$, so that the deep UV limit of $\Lambda$ then becomes, allowing $G_N(k)$
to run,
%\begin{equation}
%\begin{split}
$\Lambda(k) \operatornamewithlimits{\longrightarrow}_{k^2\rightarrow \infty} k^2\lambda_*,\;
\lambda_* =-\frac{c_{2,eff}}{2880}\sum_{j}(-1)^{F_j}n_j/\rho_j^2
\cong 0.0817$
%\end{split}
%\end{equation} 
where $F_j$ is the fermion number of $j$, $n_j$ is the effective
number of degrees of freedom of $j$ and $\rho_j=\rho(\lambda_c(m_j))$.
%, and $\rho_{avg}$ is the average
%value of $\rho$ defined as 
%\begin{equation}
%\rho_{avg}=\frac{(\sum_jn_j)(\sum_{j}(-1)^{F_j}n_j)}{c_{2,eff}\sum_{j}(-1)^{F_j}n_j/\rho_j^2},\;\; \rho_j=\rho(\lambda_c(m_j)).
%\end{equation} 
We note that
% again that $\lambda_*$ is free of threshold/cut-off effects and is
%a pure prediction of our known world -- 
$\lambda_*$ would vanish
in an exactly supersymmetric theory.\par
For reference, the UV fixed-point calculated here, 
$(g_*,\lambda_*)\cong (0.0442,0.0817)$, can be compared with the estimates
$(g_*,\lambda_*)\approx (0.27,0.36)$
in Refs.~\cite{reuter1,reuter2}.
%See Refs.~\cite{bw1} for more discussion of this comparison. 
In making this comparison, one must keep in mind that 
the analysis in Refs.~\cite{reuter1,reuter2} did not include
the specific SM matter action and that there is definitely cut-off function
sensitivity to the results in the latter analyses. What is important
is that the qualitative results that $g_*$ and $\lambda_*$ are 
both positive and are less than 1 in size 
%with $\lambda_*>g_*$
are true of our results as well.
See Refs.~\cite{bw1} for further discussion of the relationship between
our $\{g_*,\;\lambda_*\}$ predictions and those in Refs.~\cite{reuter1,reuter2}.\par

\section{\bf An Estimate of $\Lambda$ and Constraints on SUSY GUTS}

The results here, taken together with those in Refs.~\cite{reuter1,reuter2}, allow us to estimate the value of $\Lambda$ today. We take the normal-ordered form of Einstein's equation 
\begin{equation}
:G_{\mu\nu}:+\Lambda :g_{\mu\nu}:=-\kappa^2 :T_{\mu\nu}: .
\label{eineq2}
\end{equation}
The coherent state representation of the thermal density matrix then gives
the Einstein equation in the form of thermally averaged quantities with
$\Lambda$ given by our result in (\ref{lambscalar}) summed over 
the degrees of freedom as specified above in lowest order. In Ref.~\cite{reuter2}, it is argued that the Planck scale cosmology description of inflation gives the transition time between the Planck regime and the classical Friedmann-Robertson-Walker(FRW) regime as $t_{tr}\sim 25 t_{Pl}$. (We discuss in Refs.~\cite{bw1}on the uncertainty of this choice of $t_{tr}$.)
We thus start with the quantity
%\begin{equation}
%\begin{split}
$\rho_\Lambda(t_{tr}) \equiv\frac{\Lambda(t_{tr})}{8\pi G_N(t_{tr})}
         =\frac{-M_{Pl}^4(k_{tr})}{64}\sum_j\frac{(-1)^Fn_j}{\rho_j^2}$
%\end{split}
%\label{eq-rho-lambda}
%\end{equation}
and employ the arguments in Refs.~\cite{branch-zap} ($t_{eq}$ is the time of matter-radiation equality) to get the 
first principles field theoretic estimate
\begin{equation}
\begin{split}
\rho_\Lambda(t_0)&\cong \frac{-M_{Pl}^4(1+c_{2,eff}k_{tr}^2/(360\pi M_{Pl}^2))^2}{64}\sum_j\frac{(-1)^Fn_j}{\rho_j^2}\cr
          &\qquad\quad \times \frac{t_{tr}^2}{t_{eq}^2} \times (\frac{t_{eq}^{2/3}}{t_0^{2/3}})^3\cr
%         & \qquad\qquad {\Color{Magenta}\Updownarrow} \qquad\qquad\qquad {\Color{Brown}\Updownarrow} \cr
%         &\qquad {\Color{Magenta}\text{Rad. Dom.}} \qquad {\Color{Brown}\text{Mat. Dom.}}\qquad\qquad {\Color{PineGreen}\Rightarrow}\cr
    &\cong \frac{-M_{Pl}^2(1.0362)^2(-9.194\times 10^{-3})}{64}\frac{(25)^2}{t_0^2}\cr
   &\cong (2.4\times 10^{-3}eV)^4.
\end{split}
\label{eq-rho-expt}
\end{equation}
where we take the age of the universe to be $t_0\cong 13.7\times 10^9$ yrs. 
In the latter estimate, the first factor in the second line comes from the period from
$t_{tr}$ to $t_{eq}$ which is radiation dominated and the second factor
comes from the period from $t_{eq}$ to $t_0$ which is matter dominated
\footnote{The method of the operator field forces the vacuum energies to follow the same scaling as the non-vacuum excitations.}.
This estimate should be compared with the experimental result~\cite{pdg2008}\footnote{See also Ref.~\cite{sola2} for an analysis that suggests 
a value for $\rho_\Lambda(t_0)$ that is qualitatively similar to this experimental result.} 
$\rho_\Lambda(t_0)|_{\text{expt}}\cong ((2.37\pm 0.05)\times 10^{-3}eV)^4$. 
\par
To sum up, we believe our estimate 
of $\rho_\Lambda(t_0)$
represents some amount of progress in
the long effort to understand its observed value  
in quantum field theory. Evidently, the estimate is not a precision prediction,
as hitherto unseen degrees of freedom, such as a high scale GUT theory, 
may exist that have not been included in the calculation.\par
We now comment on what would happen to our estimate if there would be a GUT theory at high scale. As is well-known, the main
viable approaches involve susy GUT's and for definiteness,
we will use the susy SO(10) GUT model in Ref.~\cite{ravi-1}
to illustrate how such theory might affect our estimate of $\Lambda$.
In this model, the break-down of the GUT gauge symmetry to the 
low energy gauge symmetry occurs with an intermediate stage with gauge group
$SU_{2L}\times SU_{2R}\times U_1\times SU(3)^c$ where the final break-down to the Standard Model~\cite{gsw,qcd} gauge group, $SU_{2L}\times U_1\times SU(3)^c$, occurs at a scale $M_R\gtrsim 2TeV$ while the breakdown of global susy occurs at the (EW) scale $M_S$ which satisfies $M_R > M_S$. For our purposes
the key observation is that susy multiplets do not contribute to our formula
for $\rho_\Lambda(t_{tr})$ when susy is not broken -- there is exact cancellation between fermions and bosons in a given degenerate susy multiplet. Thus only the the broken susy multiplets can contribute. In the model at hand, these are just the multiplets associated with the known SM particles and the extra Higgs multiplet required by susy in the MSSM~\cite{haber}.
In view of recent LHC results~\cite{lhc-susy}, we take for illustration the values $M_R\cong 4 M_S\sim 2.0{\text{TeV}}$ and set the following susy partner values:
%\begin{equation}
%\begin{split}
$m_{\tilde{g}}\cong 1.5(10){\text{TeV}},\;
m_{\tilde{G}}\cong 1.5{\text{TeV}},\;
m_{\tilde{q}}\cong 1.0{\text{TeV}},\;
m_{\tilde{\ell}}\cong 0.5{\text{TeV}},\;
m_{\tilde{\chi}^0_i}\cong\begin{cases} &0.4{\text{TeV}},\;i=1\\
                                        & 0.5{\text{TeV}},\; i=2,3,4
                    \end{cases},\;
m_{\tilde{\chi}^{\pm}_i}\cong  0.5{\text{TeV}},\; i=1,2,\;
m_{S} = .5{\text{TeV}},\; S=A^0,\; H^{\pm},\; H_2,$
%\end{split}
%\end{equation}  
where we use a standard notation for the susy partners of the known quarks($q\leftrightarrow \tilde{q}$), leptons($\ell\leftrightarrow \tilde{\ell}$) and gluons($G\leftrightarrow \tilde{G}$), and the EW gauge and Higgs bosons($\gamma,\; Z^0,\; W^{\pm},\;H,$
$A^0,\;H^{\pm},\;H_2  \leftrightarrow \tilde{\chi}$)  with the extra Higgs particles denoted as usual~\cite{haber} by $A^0$(pseudo-scalar), $H^{\pm}$(charged) and $H_2$(heavy scalar). $\tilde{g}$ is the gravitino, for which we show two examples of its mass for illustration. 
These particles then generate the extra contribution 
%\begin{equation}
%\begin{split}
$\Delta W_{\rho,\text{GUT}}=\sum_{j\in \{\text{MSSM low energy susy partners}\}}\frac{(-1)^Fn_j}{\rho_j^2}
          \cong 1.13(1.12)\times 10^{-2}$
%\end{split} 
%\end{equation}
to the factor $W_\rho\equiv \sum_j\frac{(-1)^Fn_j}{\rho_j^2}$ on the RHS of 
our equation for $\rho_\Lambda(t_{tr})$ for the two respective values of $m_{\tilde g}$ called out by the parentheses. The corresponding values of $\rho_\Lambda$ are $-(1.67\times 10^{-3}\text{eV})^4(-(1.65\times 10^{-3}\text{eV})^4)$, respectively. The sign of these results would appear to put them in conflict with the positive observed value quoted above by many standard deviations, even when we allow for the considerable uncertainty in the various other factors multiplying $W_\rho$ in our formula for $\rho_\Lambda(t_{tr})$, all of which are positive in our framework. This may be alleviated either by adding new particles to the model, approach (A), or by allowing a soft susy breaking mass term for the gravitino that resides near the GUT scale
$M_{GUT}$, which is $\sim 4\times 10^{16} GeV$ here~\cite{ravi-1}, approach (B). In approach (A), we double the number of quarks and leptons, but we invert
the mass hierarchy between susy partners, so that the new squarks and sleptons are lighter than the new quarks and leptons. This can work as long as
as we increase $M_R,\; M_S$ so that we have the new quarks and leptons
at $M_{\text{High}}\sim 3.4(3.3)\times 10^3\text{TeV}$ while leaving their partners at $M_{\text{Low}}\sim .5{\text{TeV}}$. For approach (B), the mass of the gravitino soft breaking term should be set to
$m_{\tilde{g}}\sim 2.3\times 10^{15}{\text{GeV}}$. More generally, our 
estimate in (\ref{eq-rho-expt}) can be used
as a constraint of general susy GUT models and we hope to explore such in more detail elsewhere. 
%This admittedly 
%limited discussion of susy GUT effects highlights what
%one can expect for the impact on our estimate in (\ref{eq-rho-expt}) from higher mass
%scale physics.
\par 
Realistically, as we explain in Refs.~\cite{bw1},
we stress that we actually do not know the precise value of $t_{tr}$ at this point to better than a couple of orders of magnitude which translate to an uncertainty at the level of $10^4$ on
our estimate of $\rho_\Lambda$. We caution the reader to keep this in mind.
\par
There is one further important matter that we have not mentioned: the effect of the various spontaneous symmetry vacuum energies on our 
$\rho_{\Lambda}$ estimate. From the standard methods we know for example that the energy of the broken vacuum for the EW case contributes an amount of order $M_W^4$ to $\rho_\Lambda$. If we consider the GUT 
symmetry breaking we expect an analogous contribution from spontaneous symmetry breaking of order $M_{GUT}^4$. When compared to the RHS of 
our equation for $\rho_\Lambda(t_{tr})$, which is $\sim (-(1.0362)^2W_\rho/64)M_{Pl}^4\simeq \frac{10^{-2}}{64}M_{Pl}^4$, we see that adding these effects thereto would make relative changes in 
our results at the level of 
$\frac{64}{10^{-2}}\frac{M_W^4}{M_{Pl}^4}\cong 1\times 10^{-65} $ and $\frac{64}{10^{-2}}\frac{M_{GUT}^4}{M_{Pl}^4}\cong 7\times 10^{-7}$, respectively, where we use our value of $M_{GUT}$ given above in the latter 
evaluation for definiteness. 
We ignore such small effects here.
\par
We want however to stress again that 
the model Planck scale cosmology of Bonanno and Reuter
which we use is just that, a model. More work needs to be done to remove 
from it the type of uncertainties which we just elaborated in our estimate of $\Lambda$. We thank Profs. L. Alvarez-Gaume and W. Hollik for the support and kind
hospitality of the CERN TH Division and the Werner-Heisenberg-Institut, MPI, Munich, respectively, where a part of this work was done.\par
\par
%\section*{Note Added:}
%Here, we point out for clarity that in computing $\Lambda$
%in the Planck regime the assumption of $K=0$ is presumed 
%as that is the only case for which the Bonanno-Reuter Planck scale
%cosmology has been shown to allow a smooth connection from
%the Planck regime for times near or earlier than the Planck time
%to the semi-classical FRW regime for times after $t_{tr}$.
%For $K=0$, by definition, equal time slices are flat 3-spaces, exactly
%as we have employed in the vacuum states used to compute 
%the zero-point energies that comprise $\Lambda$. Thus the results
%in Sections 3 and 4 are fully self-consistent.

\end{document}